\documentclass[journal]{IEEEtran}
\usepackage{stmaryrd}
\usepackage{amsfonts}
\usepackage{amsmath}
\usepackage{amssymb}
\usepackage{cite}
\usepackage{graphicx}
\usepackage{subfigure}
\usepackage{amsthm}
\usepackage{algorithmic}
\usepackage{algorithm}
\usepackage{stfloats}

\newtheorem{Theorem}{Theorem}
\newtheorem{Lemma}{Lemma}

\newtheorem{Corollary}[Lemma]{Corollary}
\newtheorem{Example}{Example}
\def\QED{\IEEEQED\vspace{0.1in}}
\newcommand{\nchoosek}[2]{\binom{#1}{#2}}
\begin{document}
%
\title{Linear Transformations for Randomness Extraction}

\author{Hongchao~Zhou,
        and~Jehoshua~Bruck,~\IEEEmembership{Fellow,~IEEE}
\thanks{
This work was supported in part by the NSF Expeditions in Computing
Program under grant CCF-0832824. This paper was presented in part at IEEE International Symposium on Information Theory (ISIT), St. Petersburg, Russia, August 2011.
}
\thanks{Hongchao Zhou and Jehoshua~Bruck are with the Department of Electrical Engineering, California Institute of Technology, Pasadena,
CA 91125, USA, e-mail: hzhou@caltech.edu; bruck@caltech.edu}
}


\maketitle%
\begin{abstract}
Information-efficient approaches for extracting randomness from imperfect sources have been extensively studied, but
simpler and faster ones are required in the high-speed applications of random number generation.
In this paper, we focus on linear constructions, namely,
applying linear transformation for randomness  extraction.
We show that linear transformations based on sparse random matrices are asymptotically optimal
to extract randomness from independent sources and bit-fixing sources, and they are efficient (may not be optimal)
to extract randomness from hidden Markov sources. Further study demonstrates the flexibility of such constructions on source models as well as their excellent information-preserving capabilities. Since linear transformations based on sparse random matrices are computationally fast and can be easy to implement using hardware like FPGAs, they are very attractive
in the high-speed applications.  In addition, we explore explicit constructions of transformation matrices.
We show that the generator matrices of primitive BCH codes are good choices, but linear transformations based on such matrices require more
computational time due to their high densities.
\end{abstract}

\begin{IEEEkeywords}
Randomness Extraction, Linear Transformations, Sparse Random Matrices.
\end{IEEEkeywords}

\IEEEpeerreviewmaketitle

\section{Introduction}

\IEEEPARstart{R}{andomness} plays an important role in many fields, including complexity theory, cryptography, information theory and optimization. There are
many randomized algorithms that are faster, more space efficient or simpler than any known deterministic algorithms \cite{Motwani95}; hence, how to
generate random numbers becomes an essential question in computer science.
Pseudo-random numbers have been studied, but they cannot perfectly simulate truly random bits or have security issues
in some applications.
 These problems motivate people to extract random bits from natural sources directly. In this paper, we study linear transformation for randomness extraction. This approach is attractive due to its computational simplicity and information efficiency. Specifically, given an input binary sequence $X$ of length $n$ generated from an imperfect source, we construct an $n\times m$ binary matrix $M$ called a transformation matrix
such that the output sequence $$Y=XM$$ is
very close to the uniform distribution on $\{0,1\}^m$.
Statistical distance \cite{Sha02} is commonly used to measure the distance between two distributions in randomness extraction.
We say $Y\in \{0,1\}^m$ is $\epsilon$-close to the uniform distribution $U_m$ on $\{0,1\}^m$ if
and only if
\begin{equation}\label{equ_intro_1}
\frac{1}{2}\sum_{y\in\{0,1\}^m}|P[Y=y]-2^{-m}|\leq \epsilon,
\end{equation}
where $\epsilon>0$ can be arbitrarily small. This condition guarantees that in any probabilistic application, if we replace truly random bits with the sequence $Y$, the additional error probability caused by the replacement is at most $\epsilon$.

The classical question in randomness extraction considers ideal sources, like biased coins or Markov chains. From such sources,
the bits extracted can be perfectly random that means independent and unbiased. It dates back to von Neumann \cite{Neumann1951} who first considered the problem of simulating an unbiased coin by using a biased coin with unknown probability. His beautiful algorithm was later
improved by Elias \cite{Elias1972} and Peres \cite{Peres1992}. In 1986, Blum  \cite{Blum1986} studied the problem of generating
random bits from a correlated source, specifically, he considered finite Markov chains. Recently, we generalized Blum's method and proposed the first known algorithm that runs in expected linear time and achieves the information-theoretic upper bound on efficiency \cite{Zhou_Markov}. Although it is known how to extract random bits optimally from biased coins or Markov chains, these models are too narrow to describe real sources that suffer noise and disturbance.

During last two decades, research has been focused on a general source model called $k$-sources \cite{Zuc90}, in which each possible sequence has probability at most $2^{-k}$ of being generated. This model can cover a very wide range of natural random sources, but it was shown that
it is impossible to derive a single function that extracts even a single bit of randomness from such a source. This observation led to
the introduction of seeded extractors, which use a small number of truly random bits as the seed (catalyst). When simulating a probabilistic algorithm, one can simply eliminate the requirement of truly random bits by enumerating all possible strings for the seed and taking a majority vote. There are a variety of very efficient constructions of seeded extractors, summarized in \cite{Nis96, Sha02, Dvir08}. Although seeded extractors are information-efficient and applicable to most natural sources, they are not computationally fast when simulating probabilistic algorithms.
Recently, there is renewed interest in designing seedless extractors, called deterministic extractors. Several specific classes of sources have been studied, including independent sources, which can be divided into several independent parts consisting of certain amounts of randomness \cite{Rao2007, Raz2005, Rao08,Barak06}; bit-fixing sources, where some bits in a binary sequence are truly random and the remaining bits are fixed \cite{Cohen89, Kamp06, Gabizon06}; and samplable sources, where the source is generated by a process that has a bounded amount of computational resources like space\cite{Trevisan00, Kamp11}.

Unlike prior works on deterministic extractors, we take both simplicity and efficiency into consideration. Simplicity is certainly an important issue; for example, it motivates the use of von Neumann's scheme  \cite{Neumann1951} in Intel's random number generator (RNG) \cite{Jun99} rather than some other more sophisticated extractors. However, von Neumann's scheme is far from optimal in its efficiency, and it only works for ideal biased coins.
Recently, in order to support future generations of hardware security in systems
operating at ultrafast bit rates, many high-speed random number generators based on chaotic semiconductor
lasers have been developed \cite{Uchida2008}. They can generate random bits at rates as high as $12.5-400$ Gbit/s \cite{Reidler2009,Kanter2010,Akizawa2012}; hence, the simplicity of post-processing
is becoming more important.
These challenges motivate us to develop extractors that can
extract randomness from natural sources in a manner that reaches the theoretical upper bound on efficiency without compromising simplicity. In particular, we focus on linear constructions; that is, we apply linear transformations for randomness extraction.

Our main contribution is to show that linear transformations based on sparse random matrices are asymptotically optimal for extracting randomness from independent sources and bit-fixing sources, and they are efficient (although not necessarily optimal) for extracting randomness from hidden Markov sources. We further show that these conclusions hold if we apply any invertible linear mapping on the sources. In fact, many natural sources for the purpose of
high-speed random number generation are qualified to fit one of the above models or their mixture, making the construction based on sparse random matrices very attractive in practical use. The resulting extractors are not seeded extractors, which consume truly random bits whenever extracting randomness. They are, in some sense, probabilistic constructions of deterministic extractors. In addition, we explore explicit constructions of transformation matrices.
We show that the generator matrices of primitive BCH codes are good choices, but linear transformations based on such matrices require more
computational time due to their high densities.

The remainder of this paper is organized as follows. In Section \ref{lin_section_lineartransformation} we give an intuitive overview of linear transformations
for randomness extraction and present some general properties.
In Section \ref{lin_section_source}, we introduce the source models to be addressed in this paper and briefly describe our main results.
The detailed discussions for each source model, including independent sources, hidden Markov sources, bit-fixing sources and linear-subspace sources, are given
in Section \ref{lin_section_coin}, Section \ref{lin_section_Markov}, Section \ref{lin_section_bitfixing} and Section \ref{lin_section_subspace}, respectively.
In Section \ref{lin_section_stream}, we briefly describe implementation issues followed by concluding remarks in Section \ref{lin_section_conclusion}.

\section{Linear Transformations}
\label{lin_section_lineartransformation}

Let us start from a simple and fundamental question in random number generation: given a set of coin tosses $x_1, x_2, ...,x_n$ with
$P[x_i=1]\in [\frac{1}{2}-\delta,\frac{1}{2}+\delta]$, how can we simulate a single coin toss such that is as unbiased as possible? This question
has been well studied and it is known that binary sum operation is optimal among all the methods, i.e., we generate a bit $z$ which is
$$z=x_1+x_2+...+x_n\mod 2.$$
The following lemma shows that binary sum operation can decrease the bias of the resulting coin toss exponentially.

\begin{Lemma} \emph{\cite{Lac08}}\label{lemma1}
Let $x_1,x_2,...,x_n$ be $n$ independent bits and the bias of $x_i$ is $\delta_i$, namely,
$$\delta_i=|P[x_i=1]-\frac{1}{2}|$$
for $1\leq i\leq n$, then the bias of
$z=x_1+x_2+...+x_n \mod 2$ is upper bounded by
$$\frac{\prod_{i=1}^n (2\delta_i)}{2}.$$
\end{Lemma}

A generalization of the above question is that: given $n$ independent bits, how do we generate
$m<n$ random bits such that their statistical distance to the truly random bits
is as small as possible? One way is to divide all the $n$ independent bits
into $m$ nonoverlap groups, denoted by $S_1, S_2, ..., S_m$, such that $\bigcup_{i=1}^m S_i=\{x_1, x_2, ..., x_n\}$. For $1\leq i\leq m$,  the $i$th output bit, denoted by $y_i$, is produced by summing up the bits in $S_i$ and modulo two. However, this method is not very efficient.
By allowing overlaps between different groups, the efficiency can be significantly improved.
In this case, although we have sacrificed a little independence of the output bits, but the bias of each bit has been reduced a lot.
An equivalent way of presenting this method is to use a binary matrix, denoted by $M$, such that  $M_{ij}=1$ if and only if $x_i \in S_j$, otherwise, $M_{ij}=0$.
As a result, the output of this method is $Y=XM$ for a given input sequence $X$. This is an intuitive understanding why linear transformations
can be used in random extraction from weak random sources, in particular, from independent sources.

In this paper, we study linear transformations for extracting randomness from a few types of random sources.
Given a source $X\in \{0,1\}^n$, we design a transformation matrix $M$ such that the output $Y=XM$ is arbitrarily close to truly random bits. Here, we use the statistical distance between $Y$ and the uniform distribution over
$\{0,1\}^m$ to measure the goodness of the output sequence $Y$, defined by
\begin{equation}\label{equ_coin_1}
\rho(Y)=\frac{1}{2}\sum_{y\in\{0,1\}^m}|P[Y=y]-2^{-m}|.
\end{equation}
It indicates the maximum error probability introduced by replacing truly random bits with the sequence
$Y$ in any randomized algorithm.

Given a random source $X$ and a matrix $M$, the following lemma shows an upper bound of $\rho(XM)$.

\begin{Lemma}\label{lemma3}
Let $X=x_1x_2...x_n$ be a binary sequence generated from an arbitrary random source and let $M$ be an $n\times m$ binary matrix with $m\leq n$. Then given $Y=XM$,
we have
$$\rho(Y)\leq \sum_{u\in \{0,1\}^m, u\neq 0}  |P_X[XMu^T=1]-\frac{1}{2}|.$$
\end{Lemma}

\proof
Similar as the idea in \cite{Lac08}, for all $y\in \{0,1\}^m$, we define function $h$ as $h(y)=P(Y=y)$. For this function, its Fourier transform is denoted by $F_h$, then
\begin{equation*}\label{lin_equ_coin_1}
\forall y\in \{0,1\}^m, h(y)=2^{-m}\sum_{u\in\{0,1\}^m} F_h(u)(-1)^{y\cdot u},
\end{equation*}
and
$$\forall u\in \{0,1\}^m, F_h(u)=\sum_{y\in\{0,1\}^m} h(y)(-1)^{y\cdot u}.$$
When $u=0$, we have
\begin{equation*}|F_h(u)|=\sum_{y\in\{0,1\}^m} h(y)=1.\end{equation*}
When $u\neq 0$, we have
\begin{eqnarray}
  |F_h(u)| &=& |\sum_{y\in\{0,1\}^m} h(y)(-1)^{y\cdot u}| \nonumber\\
   &=& |\sum_{y\cdot u=0} h(y) - \sum_{y\cdot u=1} h(y)|\nonumber\\
   &=& |1-2\sum_{y\cdot u=1} h(y)|\nonumber \\
   &=& 2|P[XMu^T=1]-\frac{1}{2}|. \label{equ_coin_6}
\end{eqnarray}

Substituting (\ref{equ_coin_6}) into (\ref{equ_coin_1}) leads to
\begin{eqnarray}
\rho(Y)
&=& \frac{1}{2}\sum_{y\in\{0,1\}^m} |2^{-m}\sum_{u\in\{0,1\}^m} F_h(u)(-1)^{y\cdot u}-2^{-m}|\nonumber\\
&\leq & \frac{1}{2} \sum_{y\in\{0,1\}^m} 2^{-m} \sum_{ u\neq 0} |F_h(u)|\nonumber\\
&= & \frac{1}{2} \sum_{ u\neq 0} |F_h(u)|\nonumber\\
&\leq & \sum_{ u\neq 0}|P[XMu^T=1]-\frac{1}{2}|.
\end{eqnarray}
This completes the proof.
\hfill\QED

There are some related works focusing on the constructions of linear transformations for the purpose of randomness extraction.
In \cite{Lac09}, Lacharme studied linear correctors, and his goal is to generate a random sequence $Y$ of length $m$ such that
$$\max_{y\in\{0,1\}^m}|P[Y=y]-2^{-m}|\leq \epsilon$$
for a specified small constant $\epsilon$. At almost the same time as our work, in \cite{Abbe2011}, Abbe uses polar codes to construct deterministic
extractors. His idea is that given an independent sequence $X$ and let $X'=XG_n$ with $G_n=[\begin{array}{cc}
                                                                                                1 & 0 \\
                                                                                                1 & 1
                                                                                              \end{array}
]^{\bigoplus\log_2 n}$, then a subset of components in $X'$ are roughly i.i.d. uniform and the remaining components are roughly deterministic.
It was proved that this approach can generate a random sequence $Y$ of length $m$ and with entropy at least $m(1-\epsilon)$. In both of the works above, the random bits generated are `weaker' than the requirement of statistical distance. For instance, let $Y$ be a random sequence of length $m$, and assume $P[Y=y]$ with $y\in \{0,1\}^m$ is either $2^{-(m-1)}$ or $0$. In this case, as $m\rightarrow \infty$, we have
$$\max_{y\in\{0,1\}^m}|P[Y=y]-2^{-m}|\rightarrow 0;$$
$$1-\frac{H(Y)}{m}=\frac{1}{m}\rightarrow 0.$$
That means this sequence $Y$ satisfies the requirement of randomness in both of the works. But if we consider the statistical distance of $Y$ to the uniform distribution on $\{0,1\}^m$, it is
$$\rho(Y)=\frac{1}{2}\sum_{y\in\{0,1\}^m}|P[Y=y]-2^{-m}|=\frac{1}{4}.$$
That does not satisfy our requirement of randomness in the sense of statistical distance. From this point, we generate random bits with higher requirement on quality than the above works.

In the rest of this paper, we investigate those random sources on $\{0,1\}^n$ such that by applying linear transformations we can get a random sequence $Y$ with $\rho(Y)\rightarrow 0$ as $n\rightarrow \infty$.

\section{Source Models and Main Results}
\label{lin_section_source}

In this section, we introduce a few types of random sources including independent sources, hidden Markov sources,
bit-fixing sources, and linear-subspace sources, and we summarize our main results for each type of sources.
Two constructions of linear transformations will be presented and analyzed. The first construction
is based on sparse random matrices. We say a random matrix with each entry being one with probability $p$ is sparse if and only if $p$ is small and $p=w(\frac{\log n}{n})$ that means
$p>\frac{k}{\log n}{n}$ for any fixed $k>0$ when the source length $n\rightarrow \infty$.
The second construction is explicit -- it is based on the generator matrices of linear codes with binomial weight distributions.
The drawback of this construction is that it requires more computations than the first one.

Given a source $X$, let $H_{\min}(X)$ denote its min-entropy, defined by
\begin{equation}\label{equ_intro_2}H_{\min}(X)=\min_{x\in \{0,1\}^n} \log_2 \frac{1}{P[X=x]}.\end{equation}
For many sources, such as independent sources and bit-fixing sources, the number of randomness that can be extracted using deterministic extractors is upper bounded by the min-entropy of the source asymptotically. Note that this is not always true for some special sources when the input sequence is infinitely long. For example, we consider a source on $\{0,1\}^n$ such that there is one assignment with probability $2^{-\frac{n}{2}}$ and
all the other assignments have probability either $2^{-n}$ or $0$. For this source, its min-entropy is $\frac{n}{2}$, but
as $n\rightarrow\infty$, this source itself is arbitrarily close to the uniform distribution on $\{0,1\}^n$.

\subsection{Independent Sources}

Independent sources, where the bits generated are independent of each other, have been studied by Santha and Vazirani \cite{Santha86}, Varirani \cite{Vaz87}, P. Lacharme \cite{Lac09}, etc.
We consider a general model of independent sources, namely, let $X=x_1x_2...x_n\in \{0,1\}^n$ be a binary sequence generated from such a source, then $x_1,x_2,...,x_n$ are independent of each other, and all their probabilities are unknown and may be different. We assume that this source
contains a certain amount of randomness, i.e., its min-entropy $H_{\min}(X)$ is known.

\begin{Theorem}\label{theorem1_1}
Let $X=x_1x_2...x_n\in\{0,1\}^n$ be an independent sequence  and let $M$ be an $n\times m$ binary random matrix in which each entry
is $1$ with probability $p= w(\frac{\log n}{n})\leq \frac{1}{2}$.
Assume $Y=XM$.
If $\frac{m}{H_{\min}(X)}<1$, as $n\rightarrow\infty$, $\rho(Y)$ converges to $0$ in probability, i.e.,
$$\rho(Y) {\quad \buildrel p \over \rightarrow \quad } 0.$$
\end{Theorem}

It shows that linear transformations based on sparse random matrices are asymptotically optimal for extracting
randomness from independent sources. To consider explicit constructions, we focus on a type of
independent sources $X=x_1x_2...x_n\in \{0,1\}^n$ such that the probability of $x_i$ for all $1\leq i\leq n$ is slightly unpredictable, i.e.,
$$p_i=P[x_i=1]\in [\frac{1}{2}-\frac{e}{2},\frac{1}{2}+\frac{e}{2}],$$
with a constant $e$. For such a source, it is possible to have min-entropy $n\log_2\frac{2}{1+e}$. The following result
shows that we can have an explicit construction that can extract as many as $n\log_2\frac{2}{1+e}$ random bits from $X$ asymptotically.

\begin{Theorem}\label{theorem1_2}
Let $C$ be a linear code with dimension $m$ and codeword length $n$. Assume
its weight distribution is binomial and its generator matrix is $G$. Let $X=x_1x_2...x_n\in \{0,1\}^n$ be an independent source such that $P[x_i=1]\in[\frac{1}{2}-e/2,\frac{1}{2}+e/2]$ for all $1\leq i\leq n$, and let $Y=XG^T$. If $\frac{m}{n\log_2\frac{2}{1+e}}<1$, as $n\rightarrow\infty$, we have
$$\rho(Y) \rightarrow 0.$$
\end{Theorem}

This result shows that if we can construct a linear code with binomial weight distribution, it can extract as many as
$n\log_2\frac{2}{1+e}$ random bits asymptotically.
It is known that primitive BCH codes have approximately binomial weight distribution. Hence, they are good candidates for extracting
randomness from independent sources with bounded bias.

\subsection{Hidden Markov Sources}

A more-useful but less-studied model is a hidden Markov source. It is a good description of many natural sources for the purpose of high-speed random number generation, such as those based on thermal noise or clock drift. Given a binary sequence $X=x_1x_2...x_n\in \{0,1\}^n$ produced by such a source, we let $\theta_i$ be the complete information about the system at time $i$ with $1\leq i\leq n$. Examples of this system information include
 the value of the noise signal, the temperature, the environmental effects, the bit generated at time $i$, etc. So the bit generated at time $i$, i.e., $x_i$, is just a function of $\theta_i$. We say that this source has the hidden Markov property  if and only if for all $1< i\leq n$,
$$P[x_i|\theta_{i-1},x_{i-1}, x_{i-2},...,x_1]=P[x_i|\theta_{i-1}].$$
That means the bit generated at time $i$ only depends on the complete system information at time $i-1$.

To analyze the performance of linear transformations on hidden Markov sources, we assume that the external noise of the sources is bounded, hence, we assume that for any three time points $1\leq i_1<i_2<i_3<n$,
\begin{equation}\label{equ_coin_3}P[x_{i_2}=1|\theta_{i_1},\theta_{i_3}]\in [\frac{1}{2}-\frac{e}{2}, \frac{1}{2}+\frac{e}{2}]
\end{equation}
with a constant $e$.

\begin{Theorem}\label{theorem2_1}
Let $X=x_1x_2...x_n$ be a binary sequence generated from a hidden Markov source described above.
Let $M$ be an $n\times m$ binary random matrix in which the probability of each entry being
$1$ is $p= w(\frac{\log n}{n})\leq \frac{1}{2}$.
Assume $Y=XM$.
If $\frac{m}{n\log_2\frac{2}{1+\sqrt{e}}}<1$, as $n$ becomes large enough, we have that $\rho(Y)$ converges to $0$ in probability, i.e.,
$$\rho(Y) {\quad \buildrel p \over \rightarrow \quad } 0.$$
\end{Theorem}

The following theorem implies that we can also use the generator matrices of primitive BCH codes for extracting randomness from hidden Markov sources, due to their approximately binomial weight distributions.

\begin{Theorem}\label{theorem2_2}
Let $C$ be a linear binary code with dimension $m$ and codeword length $n$. Assume
its weight distribution is binomial and its generator matrix is $G$. Let $X=x_1x_2...x_n$ be a binary sequence generated from a hidden Markov source described above, and let $Y=XG^T$. If $\frac{m}{n\log_2\frac{2}{1+\sqrt{e}}}<1$, as $n\rightarrow\infty$, we have
$$\rho(Y)\rightarrow 0.$$
\end{Theorem}

Although our constructions of linear transformations are not able to extract randomness optimally from hidden Markov sources, they have good capabilities of tolerating local correlations. The gap between their
information efficiency and the optimality is reasonable small for hidden Markov sources, especially considering their constructive simplicity and the fact that most of physical sources for high-speed random number generation are roughly independent and with a very small amount of correlations.

\subsection{Bit-Fixing Sources}

Bit-fixing sources were first
studied by Cohen and Wigderson \cite{Cohen89}.
In an oblivious bit-fixing source $X$ of length $n$,
$k$ bits in $X$ are unbiased and independent, and the remaining $n-k$ bits are fixed.
We also have nonoblivious bit-fixing sources, in which
the remaining $n-k$ bits linearly depend on the $k$ independent and unbiased bits. Such sources were
originally studied in the context of
collective coin flipping \cite{Ben90}. Here, we say a bit-fixing source for the general nonoblivious case.

\begin{Theorem}\label{theorem3_1}
Let $X=x_1x_2...x_n\in \{0,1\}^n$ be a bit-fixing source in which
$k$ bits are unbiased and independent.  Let $M$ be an $n\times m$ binary random matrix in which the probability for each entry being $1$
is $p= w(\frac{\log n}{n})\leq \frac{1}{2}$.
Assume $Y=XM$.
If $\frac{m}{k}<1$, as $n$ becomes large enough, we have that $\rho(Y)=0$ with almost probability $1$, i.e.,
$$P_M[\rho(Y)= 0]\rightarrow 1.$$
\end{Theorem}

So sparse random matrices are asymptotically optimal to extract randomness from bit-fixing sources. Unfortunately, for bit-fixing sources, it is possible to find an efficient and explicit construction of linear transformations.

\subsection{Linear-Subspace Sources}

We generalize the sources described above in the following way: Assume $X\in \{0,1\}^n$ is a raw sequence that can be written as $ZA$, where $Z\in\{0,1\}^k$ with $k<n$ is an independent sequence or a hidden Markov sequence, and $A$ is an
$k\times n$ unknown matrix with full rank, i.e., it is an invertible matrix.
Instances of such sources include
sparse images studied in compressive sensing. We call such sources
as linear-subspace sources, namely, they are obtained by mapping simpler sources into a subspace of higher dimensions.
We demonstrate that linear transforms based on sparse random matrices can work on linear-subspace sources,
and any linear invertible operation on the sources does not affect the asymptotic performance.
Specifically, we have the following theorem.

\begin{Theorem}\label{theorem4_1}
Let $X=x_1x_2...x_n\in \{0,1\}^n$ be a source such that $X=ZA$ in which $Z$ is an independent sequence and $A$ is an unknown $k \times n$ full-rank matrix. Let $M$ be an $n\times m$ random matrix such that each entry of $M$
is $1$ with probability $p= w(\frac{\log n}{n})\leq \frac{1}{2}$.
Assume $Y=XM$.
If $\frac{m}{H_{\min}(X)}<1$, as $n\rightarrow\infty$,
$\rho(Y)$ converges to $0$ in probability, i.e.,
$$\rho(Y) {\quad \buildrel p \over \rightarrow \quad } 0.$$
\end{Theorem}

A similar result holds if $Z$ is a hidden Markov sequence. In this case, we only need to
replace $H_{\min}(X)$ with $k\log_2\frac{1}{1+\sqrt{e}}$, where $k$ is the length of $Z$ and $e$ is defined in Equ.~(\ref{equ_coin_3}).

\subsection{Comments}
Compared to $k$-sources, the models that we study in this paper are more specific. Perhaps, they are not perfect to describe some sources like users' operating behaviors or English articles. But for most natural sources that are used for building high-speed random number generators, they are
very good descriptions. Based on these models, we can explore simpler and more practical algorithms than those designed for general $k$-sources.
In the following sections, we will present our  technical results in detail for different types of sources respectively.

\section{Independent Sources}
\label{lin_section_coin}

In this section, we study a general independent source $X=x_1x_2...x_n\in \{0,1\}^n$, in which all the bits $x_1,x_2,...,x_n$
are independent of each other and the probability of $x_i$ with $1\leq i\leq n$ can be arbitrary value, i.e., $p_i\in [0,1]$.
We can consider this source as a biased coin with the existence of external adversaries.

\begin{Lemma}\label{lin_theorem5} Given a deterministic extractor $f:\{0,1\}^n\rightarrow\{0,1\}^m$, as $n\rightarrow\infty$, we have $\rho(f(X))\rightarrow 0$ for an arbitrary independent source $X$ only if
 $$\frac{m}{H_{\min}(X)}\leq 1,$$
where $H_{\min}(X)$ is the min-entropy of $X$.
\end{Lemma}

\proof  To prove this theorem, we only need to consider a source $X=x_1x_2...x_n\in \{0,1\}^n$ such that
$$P[x_i=1]=\frac{1}{2}, \forall 1\leq i\leq H_{\min}(X),$$
and
$$P[x_i=1]=0, \forall H_{\min}(X)<i\leq n.$$

From such a source $X$, if $m>H_{min}(X)$, it is easy to see that
$\rho(f(X))>0$ for all $n>0$.
\hfill\QED

Let us first consider a simple random matrix in which each entry is $1$ or $0$ with probability $1/2$ that we call a uniform random matrix. Given an independent input sequence $X\in \{0,1\}^n$ and an $n\times m$ uniform random matrix $M$, let $Y=MX\in \{0,1\}^m$ be the output sequence. The following lemma provides the upper bound of $E[\rho(Y)]$.

\begin{Lemma}\label{lin_theorem1}
Let $X=x_1x_2...x_n$ be an independent sequence and $M$ be an $n\times m$ uniform random matrix.
Then given $Y=XM$, we have
$$E_M[\rho(Y)]\leq 2^{m-H_{\min(X)-1}}.$$
\end{Lemma}

\proof
Let $p_i$ denote the probability of $x_i$ and let $\delta_i$ be
the bias of $x_i$, then $\delta_i=|p_i-\frac{1}{2}|$.

According to Lemma \ref{lemma1}, when $u\neq 0$, we have
\begin{eqnarray}
|P_X[XMu^T=1]-\frac{1}{2}| \leq \frac{\prod_{i=1}^n (2\delta_i)^{(Mu^T)_i}}{2},\label{equ_coin_4}
\end{eqnarray}
where $(Mu^T)_i$ is the $i$th element of the vector $Mu^T$.

Substituting (\ref{equ_coin_4}) into Lemma \ref{lemma3} yields
\begin{eqnarray}
\rho(Y)
&\leq & \frac{1}{2} \sum_{u\neq 0}  \prod_{i=1}^n (2\delta_i)^{(Mu^T)_i}. \label{equ_coin_5}
\end{eqnarray}

Now, we calculate the expectation of $\rho(Y)$, which is
\begin{eqnarray}
&&E_M[\rho(Y)]\\
&\leq& \frac{1}{2} E_M[\sum_{u\neq 0}\prod_{i=1}^n(2\delta_i)^{(Mu^T)_i}] \nonumber\\
&=& \frac{1}{2} \sum_{u\neq 0} \sum_{v\in\{0,1\}^n} P_M[Mu^T=v^T] \prod_{i=1}^n (2\delta_i)^{v_i}. \label{lin_equ_coin_2}
\end{eqnarray}

Since $M$ is a uniform random matrix (each entry is either $0$ or $1$ with probability $1/2$), if $u\neq 0$, $Mu^T$ is a random vector of length $n$ in which each element is $0$ or $1$ with probability $1/2$.
So for any $u\neq 0$, $$P_M[Mu^T=v^T]=2^{-n}.$$

As a result,
\begin{eqnarray*}
E_M[\rho(Y)] &\leq &  2^{m-n-1}  \sum_{v\in\{0,1\}^n} \prod_{i=1}^n (2\delta_i)^{v_i}\\
&=& 2^{m-1}\prod_{i=1}^n (\frac{1}{2}+\delta_i).
\end{eqnarray*}

For the independent sequence $X$, its min-entropy can be written as
\begin{eqnarray*}
H_{\min}(X)&=&\log_2\frac{1}{\prod_{i=1}^n\max(p_i,1-p_i)}\\
&=& \log_2\frac{1}{\prod_{i=1}^n (\frac{1}{2}+\delta_i)}.
\end{eqnarray*}

So
$$E_M[\rho(Y)]\leq 2^{m-H_{\min}(X)-1}.$$

This completes the proof.
\hfill\QED

\begin{Example} Let us consider an independent source $X=x_1x_2...x_{512}\in \{0,1\}^{512}$ in which
$$p_i\in [\frac{1}{2}-\frac{i}{1024}, \frac{1}{2}+\frac{i}{1024}]$$
for all $1\leq i\leq 512$.

For this source, its min-entropy is
$$H_{\min}(X)\geq -\sum_{i=1}^{512}\log_2(\frac{1}{2}+\frac{i}{1024})=226.16.$$

If we use a $512\times 180$ random matrix in which each entry is $0$ or $1$ with probability $1/2$, then according to the above lemma,
$$E[\rho(Y)]\leq 2^{-47.16}\leq 6.4\times 10^{-15}.$$
That means that the output sequence is very close to the uniform distribution in the sense of statistical distance.
\end{Example}

When $n$ is large enough, we have the following corollary, showing that uniform random matrices are capable to extract as many as $H_{\min}(X)$ random bits from an independent source $X$ asymptotically with almost probability one. Since $H_{\min}(X)$ is the theoretical upper bound, such an extractor is asymptotically optimal on efficiency.

\begin{Corollary}\label{Corollary1}
Let $X\in\{0,1\}^n$ be an independent sequence and let $M$ be an $n\times m$ uniform random matrix.
Assume $Y=XM$.
If $\frac{m}{H_{\min}(X)}<1$, as $n\rightarrow\infty$, $\rho(Y)$ converges to $0$ in probability, i.e.,
$$\rho(Y) {\quad \buildrel p \over \rightarrow \quad } 0.$$
\end{Corollary}

The above corollary shows that when the length of the input sequence $n$ is large, we can extract random bits very efficiently from an independent source by simply constructing a uniform random matrix.
We need to distinguish this method from those of seeded extractors that use some additional random bits whenever extracting randomness. In our method, the matrix is randomly generated but the extraction itself is still deterministic, that means
we can use the same matrix to extract randomness for any number of times without reconstructing it. From this point, our method is  a `probabilistic construction of deterministic extractors'.

Although linear transformations based on uniform random matrices are very efficient for extracting randomness from independent sources,
they are not computationally fast due to the high density. It is natural to ask whether it is possible to decrease
the density of $1$s in the matrices without affecting the performance too much.
Motivated by this question, we study a sparse random matrix $M$ in which each entry is $1$ with probability
$p=w(\frac{\log n}{n})\ll \frac{1}{2}$, where $p=w(\frac{\log n}{n})$ means that $p>\frac{k\log n}{n}$ for any fixed $k$ when $n\rightarrow\infty$.
Surprisingly, such a sparse matrix has almost the same performance as that of a uniform random matrix, namely,
it can extract as many as $H_{\min}(X)$ random bits when the input sequence is long enough.

\begin{Lemma}\label{lemma4}
Let $p=w(\frac{\log n}{n})\leq \frac{1}{2}$ and
let $$f_n(p)=\sum_{j=1}^{\frac{\log\frac{1}{\epsilon}}{2p}} \nchoosek{m}{j} (\frac{1}{2}(1+(1-2p)^j))^n $$
with $\epsilon>0$ and $m=\Theta(n)$. As $n\rightarrow\infty$,
we have
$$f_n(p)\rightarrow 0.$$
\end{Lemma}

\proof Since $m=\Theta(n)$, we can write $m=cn$ with a constant $c$.

Let us introduce a function $F(j)$, defined by
\begin{eqnarray*}
F(j)&=&m^j 2^{-n}(1+(1-2p)^j)^n\\
   &=& c^jn^j 2^{-n}(1+(1-2p)^j)^n.
\end{eqnarray*}

Then
$$f_n(p)\leq \sum_{j=1}^{\frac{\log\frac{1}{\epsilon}}{2p}}F(j).$$

First, if $p=\frac{1}{2}$, as $n\rightarrow\infty$, we have
\begin{eqnarray*}
f_n(p)&\leq & \sum_{j=1}^{\frac{\log\frac{1}{\epsilon}}{2p}} c^j n^j 2^{-n}\\
&\leq & \frac{\log\frac{1}{\epsilon}}{2p} 2^{log_2(cn)\frac{\log\frac{1}{\epsilon}}{2p}}2^{-n}\\
&\leq & \frac{\log\frac{1}{\epsilon}}{2p} 2^{\frac{2n\log_2(cn)}{w(\log n)}\log\frac{1}{\epsilon}-n}\\
&=& \frac{\log\frac{1}{\epsilon}}{2p} 2^{-\Theta(n)}\\
&\rightarrow& 0.
\end{eqnarray*}

If $p<\frac{1}{2}$, we show that $F(j)$ decreases as $j$ increases for $1\leq j\leq \frac{\log \frac{1}{\epsilon}}{2p}$ when $n$ is large enough. To see this,
we show that its derivative $F'(j)<0$ when for $n\rightarrow\infty$.
\begin{eqnarray*}
&&F'(j)\\
&=& c^jn^j \log(cn) 2^{-n}(1+(1-2p)^j)^n \\
&&+ c^jn^j 2^{-n} n (1+(1-2p)^j)^{n-1} (1-2p)^j \log(1-2p)\\
&\leq& c^jn^j 2^{-n} n \log(cn) 2^{-n}(1+(1-2p)^j)^n \\
&& \times [1+\frac{ (1-2p)^j \log(1-2p)n}{2\log(cn)}].
\end{eqnarray*}

So we only need to prove that
$$1+\frac{ (1-2p)^j \log(1-2p)n}{2\log(cn)}<0$$
for $n\rightarrow\infty$.

Since $p\leq \alpha<\frac{1}{2}$ for a constant $\alpha$, we have
$$(1-2p)^{-\frac{1}{2p}}\leq \beta= (1-2\alpha)^{-\frac{1}{2\alpha}},$$
where $\beta$ is a constant.

We can also have
$$\log(1-2p)\leq -2p.$$

Hence,
\begin{eqnarray*}
&& 1+\frac{ (1-2p)^j \log(1-2p)n}{2\log(cn)}\\
&\leq& 1+\frac{ (1-2p)^{\frac{\log\frac{1}{\epsilon}}{2p}} \log(1-2p)n}{2\log(cn)} \\
&\leq&  1-\frac{ \beta^{\log\epsilon }2pn}{2\log(cn)} \\
&=& 1-\frac{\beta^{\log\epsilon }2w(\frac{\log n}{n})}{2\log(cn)}\\
&<& 0.
\end{eqnarray*}

So when $p<\frac{1}{2}$ and $n\rightarrow\infty$, $F(j)$ decreases as $j$ increases for $1\leq j\leq \frac{\log \frac{1}{\epsilon}}{2p}$.
As a result, when $n$ is large enough, we have
\begin{eqnarray*}
f_n(p) &\leq &\sum_{j=1}^{\frac{\log\frac{1}{\epsilon}}{2p}}F(j)\\
&\leq & \frac{\log\frac{1}{\epsilon}}{2p} F(1)\\
&\leq &  \frac{\log\frac{1}{\epsilon}}{2p} cn (1-p)^n\\
&\leq &   (cn)^2 (1-p)^n.
\end{eqnarray*}

Since
\begin{eqnarray*}
\log f_n(p) &\leq & 2\log c+ 2\log n +n \log (1-p)\\
&\leq & 2\log c+ 2\log n -\frac{np}{2} \\
&\rightarrow&-\infty,
\end{eqnarray*}
we can conclude that
$$f_n(p)\rightarrow 0$$ as $n\rightarrow \infty$.

This completes the proof.\hfill\QED

Based on the above lemma, we get the following theorem.

\vspace{0.1in}
\hspace{-0.15in}\textbf{Theorem \ref{theorem1_1}.} \emph{Let $X=x_1x_2...x_n\in\{0,1\}^n$ be an independent sequence  and let $M$ be an $n\times m$ binary random matrix in which each entry
is $1$ with probability $p= w(\frac{\log n}{n})\leq \frac{1}{2}$.
Assume $Y=XM$.
If $\frac{m}{H_{\min}(X)}<1$, as $n\rightarrow\infty$, $\rho(Y)$ converges to $0$ in probability, i.e.,$$\rho(Y) {\quad \buildrel p \over \rightarrow \quad } 0.$$\vspace{-0.15in}}
\proof Let us use the same denotations as above.  From Equ.~(\ref{lin_equ_coin_2}) we have
$$E_M[\rho(Y)]\leq \frac{1}{2} \sum_{u\neq 0} \sum_{v\in\{0,1\}^n} P_M[Mu^T=v^T] \prod_{i=1}^n (2\delta_i)^{v_i}.$$

Since $M$ is a random matrix in which each entry is $1$ with probability $p$, for a fixed vector $u\neq 0$ with $\|u\|=j$,
$Mu^T$ is a random vector where all the entries are independent and each entry is $1$ with probability $p_j$.
Here, according to Lemma \ref{lemma1}, we have $$p_j\in [\frac{1}{2}(1-(1-2p)^j), \frac{1}{2}(1+(1-2p)^j)].$$
There are totally $\nchoosek{m}{j}$ vectors for $u$ with $\|u\|=j$, hence, we get
\begin{eqnarray*}
&&E_M[\rho(Y)]\\
&\leq & \frac{1}{2}\sum_{j=1}^m \nchoosek{m}{j} \sum_{v\in \{0,1\}^n}(\frac{1}{2}(1+(1-2p)^j))^n \prod_{i=1}^n(2\delta_i)^{v_i}\\
&= & \frac{1}{2}\sum_{j=1}^m \nchoosek{m}{j} (1+(1-2p)^j)^n\prod_{i=1}^n (\frac{1}{2}+\delta_i).
\end{eqnarray*}

Now, we divide the upper bound of $E_M[\rho(Y)]$ into two terms. To do this, we let
$$\gamma_1=\sum_{j=1}^{\frac{\log\frac{1}{\epsilon}}{2p}} \nchoosek{m}{j} (1+(1-2p)^j)^n\prod_{i=1}^n (\frac{1}{2}+\delta_i),$$
$$\gamma_2=\sum_{j=\frac{\log\frac{1}{\epsilon}}{2p}}^m  \nchoosek{m}{j}(1+(1-2p)^j)^n\prod_{i=1}^n (\frac{1}{2}+\delta_i),$$
where $\epsilon$ can be arbitrarily small, then
$$E_M[\rho(Y)]\leq \frac{\gamma_1}{2}+\frac{\gamma_2}{2}.$$

According to Lemma \ref{lemma4}, we can get that as $n\rightarrow\infty$, if $p=w(\frac{\log n}{n})\leq \frac{1}{2}$, then
$\gamma_1\rightarrow 0$. So we only need to consider the second term, that is
$$\gamma_2\leq  \sum_{j=\frac{\log\frac{1}{\epsilon}}{2p}}^m  \nchoosek{m}{j} (1+(1-2p)^{\frac{\log\frac{1}{\epsilon}}{2p}})^n\prod_{i=1}^n (\frac{1}{2}+\delta_i).$$

Since $(1-2p)^{-\frac{1}{2p}}\geq e$, we can get
$$(1-2p)^{\frac{\log\frac{1}{\epsilon}}{2p}}\leq \epsilon.$$

As a result,
 \begin{eqnarray*}
\gamma_2&\leq & \sum_{j=\frac{\log\frac{1}{\epsilon}}{2p}}^m  \nchoosek{m}{j} (1+\epsilon)^n \prod_{i=1}^n(\frac{1}{2}+\delta_i)\\
&\leq& 2^m (1+\epsilon)^n \prod_{i=1}^n (\frac{1}{2}+\delta_i)\\
&\leq& 2^{m-n\log_2(1+\epsilon)-H_{\min}(X)}.
\end{eqnarray*}

Since $\epsilon$ can be arbitrary small, if $\frac{m}{H_{\min}(X)}<1$, as $n\rightarrow \infty$, it has
$$\gamma_2\rightarrow 0.$$

We can conclude that if
$\frac{m}{H_{\min}(X)}<1$,
$E_M[\rho(Y)]$ can be arbitrarily small as $n\rightarrow\infty$. It implies that $\rho(Y) {\quad \buildrel p \over \rightarrow \quad } 0$ as $n\rightarrow \infty$.

This completes the proof.
\hfill\QED

For practical use, we can set some constraints on each column of the sparse random matrices. For example,
we can let the number of ones in each column be a constant $k$.
We may also use pseudorandom bits instead of truly random bits.
In coding theory, many good codes are constructed based on randomly generated matrices. Such examples include LDPC (low-density parity-check) codes, network coding and compressive sensing. While these codes have very good performances, efficient decoding algorithms are needed to recover the original messages. Compared to those applications, randomness extraction is a one-way process that we do not need to reconstruct input sequences
(we also cannot do this due to the entropy loss). This feature makes linear transformations based on random matrices very attractive in the applications of randomness extraction.

In the rest of this section, we study deterministic approaches for constructing linear transformations. Here, we focus on a type of independent sources that have been studied in \cite{Santha86,Vaz87,Lac09}, and we call them independent sources with bounded bias. Let $X=x_1x_2...x_n\in \{0,1\}^n$ be an independent sequence generated
from such a source, then the probability of $x_i$ for all $1\leq i\leq n$ is slightly unpredictable, namely,
$$p_i=P[x_i=1]\in [\frac{1}{2}-\frac{e}{2},\frac{1}{2}+\frac{e}{2}]$$
for a constant $e$ with $0<e<1$.

The following theorem shows that if the weight distribution of a linear code is binomial, then the transpose of its generator matrix
is a good candidate for extracting randomness from independent sources with bounded bias.

\vspace{0.1in}
\hspace{-0.15in}\textbf{Theorem \ref{theorem1_2}.} \emph{Let $C$ be a linear code with dimension $m$ and codeword length $n$. Assume
its weight distribution is binomial and its generator matrix is $G$. Let $X=x_1x_2...x_n\in \{0,1\}^n$ be an independent source such that $P[x_i=1]\in[\frac{1}{2}-e/2,\frac{1}{2}+e/2]$ for all $1\leq i\leq n$, and let $Y=XG^T$. If $\frac{m}{n\log_2\frac{2}{1+e}}<1$, as $n\rightarrow\infty$, we have
$$\rho(Y) \rightarrow 0.$$}
\vspace{-0.15in}

\proof  Following Equ.~(\ref{equ_coin_5}) in the proof of Theorem \ref{lin_theorem1}, we get
\begin{eqnarray*}
  \rho(Y) &\leq & \frac{1}{2} \sum_{u\neq 0}  e^{w((uG)^T)}\\
   &= &\frac{1}{2}\sum_{i=1}^n 2^m\frac{{\nchoosek{n}{i}}}{2^n} e^i\\
   &\leq & 2^{m-n-1}(1+e)^n.
\end{eqnarray*}

Then it is easy to see that if $\frac{m}{n\log_2\frac{2}{1+e}}<1$, as $n\rightarrow\infty$, we have
$$\rho(Y) \rightarrow 0.$$

This completes the proof.
\hfill\QED

According to the theorem above, as $n$ becomes large enough, we can extract as many as $n\log_2(\frac{2}{1+e})$ random bits
based on the generator matrix of a linear code with binomial weight distribution. Note that the min-entropy of the source is possible to be
$$H_{\min}(X)=n\log_2(\frac{2}{1+e}),$$
which can be achieved when $p_i=\frac{1}{2}+\frac{e}{2}$ for all $1\leq i\leq n$. Hence,
this construction is as efficient as that based on random matrices, both asymptotically optimal.

It turns out that the generator matrices of primitive BCH codes are good candidates. For a primitive BCH code of length $2^k-1$, it is known that the
weight distribution of the code is approximately binomial, see theorem 21 and 23 in \cite{Macwilliams77}.
Namely, the number $b_i$ of codewords of weight $i$ is
$$b_i=a{\nchoosek{2^k-1}{i}} (1+E_i),$$
where $a$ is a constant, and the error term $E_i$ tends to zero as $k$ grows.

We see that for the uniform random matrices (with each entry being $0$ or $1$ with probability $1/2$), their weight
distributions are binomial in expectation; for sparse random matrices and primitive binary BCH codes, their
weight distributions are approximately binomial. Binomial weight distribution is one of important features for `good' matrices, based on which
one can extract randomness efficiently from independent sources.

\section{Hidden Markov Sources}
\label{lin_section_Markov}

A generalized model of an independent source are a hidden Markov source. Given a hidden Markov source $X=x_1x_2...x_n\in \{0,1\}^n$, let $\theta_i$ be the complete information about the system at time $i$ with $1\leq i\leq n$. Examples of this system information include the value of the noise signal, the temperature, the environmental effects, the bit generated at time $i$, etc. So the bit generated at time $i$, i.e., $x_i$, is just a function of $\theta_i$. We say that a source has hidden Markov property if and only if for all $1< i\leq n$,
$$P[x_i|\theta_{i-1},x_{i-1}, x_{i-2},...,x_1]=P[x_i|\theta_{i-1}].$$
That means the bit generated at time $i$ only depends on the complete system information at time $i-1$. Apparently,
such sources are good descriptions of many natural sources for the purpose of high-speed random number generation, like those based on thermal noise, avalanche noise, etc.

\begin{Example} Let us consider a weak random source based on thermal noise. By sampling the noise signal, we
get a time sequence of real numbers:
$$y_1y_2...y_n\in \mathcal{R}^n.$$
For this time sequence it has Markov property, i.e.,
$$P[y_i|y_{i-1},...,y_{1}]=P[y_i|y_{i-1}].$$
By comparing the value at each time with a fixed threshold, we get a binary sequence as the source
$$X=x_1x_2...x_n\in \{0,1\}^n,$$ such that
$x_i=\mathrm{sgn}(y_i-a)$ with a constant $a$ for all $1\leq i\leq n$.
\end{Example}

To analyze the performance of linear transformations on hidden Markov sources, we assume that the external noise of the sources is bounded, hence, we assume that for any three time points $1\leq i_1<i_2<i_3<n$,
$$P[x_{i_2}=1|\theta_{i_1},\theta_{i_3}]\in [\frac{1}{2}-\frac{e}{2}, \frac{1}{2}+\frac{e}{2}]$$
for a constant $e$.

\begin{Lemma}
Let $X=x_1x_2...x_n$ be a binary sequence generated from a hidden Markov source described above.
Let $z=x_{i_1}+...+x_{i_t}\mod 2$ for $1\leq i_1<i_2<...<i_t\leq n$ with some $t$, then we have
\begin{equation}\label{equ_markov_1}
    |P[z=1]-\frac{1}{2}|\leq \frac{e^{(t-1)/2}}{2}.
\end{equation}
\end{Lemma}

\proof
\begin{eqnarray*}
&&|P[z=1]-\frac{1}{2}|\\
&=&|\sum_{\theta_{i_1},\theta_{i_3},...}P[\theta_{i_1},\theta_{i_3},...]P[z=1|\theta_{i_1},\theta_{i_3},...]-\frac{1}{2}|\\
&\leq & \sum_{\theta_{i_1},\theta_{i_3},...} P[\theta_{i_1},\theta_{i_3},...]|P[z=1|\theta_{i_1},\theta_{i_3},...]-\frac{1}{2}|\\
&\leq & \max_{\theta_{i_1},\theta_{i_3},...} |P[x_{i_2}+x_{i_4}+...|\theta_{i_1},\theta_{i_3},...]-\frac{1}{2}|.
\end{eqnarray*}

Given $\theta_{i_1}, \theta_{i_3}, ...$,
we have $x_{i_2}, x_{i_4}, ...$ independent of each other.
So the conclusion is immediate following the statement of Lemma \ref{lemma1}.
\hfill\QED

For some hidden Markov sources, the constraint $e$ is not so strict. It is possible that there exists
a group of $\theta_{i_1},\theta_{i_3},...$ such that
$$|P[z=1|\theta_{i_1},\theta_{i_3},...]-\frac{1}{2}|>\frac{e^{(t-1)/2}}{2}.$$

In this case, we may find a typical set $S$ such that
$$P[(\theta_{i_1},\theta_{i_3},...)\in S]\rightarrow 1,$$ as the sequence becomes long enough,
and in this typical set,
$$|P[z=1|(\theta_{i_1},\theta_{i_3},...)\in S]-\frac{1}{2}|\leq \frac{e^{(t-1)/2}}{2}.$$

In this case, we can write
$$|P[z=1]-\frac{1}{2}|\leq P[(\theta_{i_1},\theta_{i_3},...)\notin S]$$
$$+\max_{(\theta_{i_1},\theta_{i_3},...)\in S} |P[z|\theta_{i_1},\theta_{i_3},...]-\frac{1}{2}|,$$
where the first term on the righthand side is ignorable.

Note that Equ.~(\ref{equ_markov_1}) can be rewritten as
$$|P[z=1]-\frac{1}{2}|\leq \frac{(\sqrt{e})^t}{2\sqrt{e}},$$
which is very similar to the result in Lemma \ref{lemma1}. If we ignore the constant term $\sqrt{e}$, the only difference between them is replacing $e$ by $\sqrt{e}$.
Based on this observation as well as the results in Section \ref{lin_section_coin} for independent sources,
we can obtain the following results for hidden Markov sources.

\begin{Lemma}
Let $X=x_1x_2...x_n$ be a binary sequence generated from a hidden Markov source described above.
Let $M$ be an $n\times m$ random matrix such that each entry of $M$ is $0$ or $1$ with probability $\frac{1}{2}$.
Then given $Y=XM$, we have
$$E_M[\rho(Y)]\leq \frac{2^{m-n-1}}{\sqrt{e}} (1+\sqrt{e})^n.$$
\end{Lemma}

So with a uniform random matrix, one can extract as many as $n\log_2\frac{2}{1+\sqrt{e}}$ random bits from a hidden Markov source. And this conclusion is also true for sparse random matrices, given by the following theorem.

\vspace{0.1in}
\hspace{-0.15in}\textbf{Theorem \ref{theorem2_1}.} \emph{Let $X=x_1x_2...x_n$ be a binary sequence generated from a hidden Markov source described above.
Let $M$ be an $n\times m$ binary random matrix in which the probability of each entry being
$1$ is $p= w(\frac{\log n}{n})\leq \frac{1}{2}$.
Assume $Y=XM$.
If $\frac{m}{n\log_2\frac{2}{1+\sqrt{e}}}<1$, as $n$ becomes large enough, we have that $\rho(Y)$ converges to $0$ in probability, i.e.,
$$\rho(Y) {\quad \buildrel p \over \rightarrow \quad } 0.$$}
\vspace{-0.15in}
\proof The proof follows the same idea for the proof of Theorem \ref{theorem1_1}.\hfill\QED

\vspace{0.1in}
\hspace{-0.15in}\textbf{Theorem \ref{theorem2_2}.} \emph{Let $C$ be a linear binary code with dimension $m$ and codeword length $n$. Assume
its weight distribution is binomial and its generator matrix is $G$. Let $X=x_1x_2...x_n$ be a binary sequence generated from a hidden Markov source described above, and let $Y=XG^T$. If $\frac{m}{n\log_2\frac{2}{1+\sqrt{e}}}<1$, as $n\rightarrow\infty$, we have
$$\rho(Y)\rightarrow 0.$$}
\vspace{-0.15in}
\proof The proof follows the same idea for the proof of Theorem \ref{theorem1_2}.\hfill\QED

These theorems show that when $n$ is large enough, we can extract as many as $n\log_2\frac{2}{1+\sqrt{e}}$ random bits from the
a hidden Markov source using linear transformations.

Let us consider an order-1 Markov source as a special instance. Assume that $X=x_1x_2...x_n$ is a binary sequence generated from this source such that each bit $x_i\in\{0,1\}$ only depends on its previous one bit, namely,
$$P[x_i=1|x_{i-1}]\in [\frac{1}{2}-\varepsilon/2,\frac{1}{2}+\varepsilon/2]$$
for a constant $\varepsilon$. Note that the transition probabilities are slightly unpredictable.

We first show that such a source can be treated as a (hidden) Markov source such that
for any $1\leq i_{j-1}\leq i_j \leq i_{j+1}\leq n$,
$$|P[x_{i_{j}}|x_{i_{j-1}},x_{i_{j+1}}]-\frac{1}{2}|\leq \frac{e}{2}$$ for a constant $e$.

According to the definition, we have
\begin{eqnarray*}
&&|P[x_{i_j}|x_{i_{j-1}}]-\frac{1}{2}|\\
&=&|\sum_{x_{i_{j-1}+1},...,x_{i_j-1}} P[x_{i_j}|x_{i_{j}-1}]...P[x_{i_{j-1}+1}|x_{i_{j-1}}]-\frac{1}{2}|\\
&\leq & \sum_{x_{i_{j-1}+1},...,x_{i_j-1}} P[x_{i_{j}-1}|x_{i_{j}-2}]...P[x_{i_{j-1}+1}|x_{i_{j-1}}]\\
&&\quad\quad\quad\quad\quad\quad \times |P[x_{i_j}|x_{i_{j}-1}]-\frac{1}{2}|\\
&\leq& \frac{\varepsilon}{2}.
\end{eqnarray*}

As a result,
\begin{eqnarray*}
&&|P[x_{i_{j}}|x_{i_{j-1}},x_{i_{j+1}}]-\frac{1}{2}|\\
&\leq & |\frac{P[x_{i_{j-1}}]P[x_{i_{j}}|x_{i_{j-1}}]P[x_{i_{j+1}}|x_{i_{j}}]}{\sum_{x_{i_{j}}} P[x_{i_{j-1}}]P[x_{i_{j}}|x_{i_{j-1}}]P[x_{i_{j+1}}|x_{i_{j}}]}-\frac{1}{2}|\\
&\leq & |\frac{(\frac{1}{2}+\frac{\varepsilon}{2})^2}{(\frac{1}{2}+\frac{\varepsilon}{2})^2+(\frac{1}{2}-\frac{\varepsilon}{2})^2}-\frac{1}{2}|\\
&= & \frac{\varepsilon}{1+\varepsilon^2}.
\end{eqnarray*}

Then, by setting $e=\frac{2\varepsilon}{1+\varepsilon^2}$, we can get
$$|P[x_{i_{j}}|x_{i_{j-1}},x_{i_{j+1}}]-\frac{1}{2}|\leq \frac{e}{2}$$
for all $1\leq i_{j-1}\leq i_j \leq i_{j+1}\leq n$.

According to the above theorems, with linear transformations,
we can extract as many as $n\log_2(\frac{2}{1+\sqrt{\frac{2\varepsilon}{1+\varepsilon^2}}})$ random bits from the above source asymptotically. In this case,
$$n\log_2(\frac{2}{1+\sqrt{\frac{2\varepsilon}{1+\varepsilon^2}}})\leq \min_{X} H_{\min}(X)=n\log_2(\frac{2}{1+\varepsilon}).$$
That means the linear transformations are not optimal for extracting randomness from order-1 Markov sources. It is true for most hidden Markov sources.
But we need to see that linear transformations have good capabilities of tolerating local correlations. The gap between their
information efficiency and the optimality is reasonable small for hidden Markov sources, especially considering their constructive simplicity.
In high-speed random number generation, the physical sources usually have relatively good quality, namely, the bits are roughly independent (with a very small amount of correlations). In this case, Linear transformation are
very efficient in extracting randomness.

\section{Bit-Fixing Sources}
\label{lin_section_bitfixing}

In this section, we consider another type of weak random sources, called bit-fixing sources, first
studied by Cohen and Wigderson \cite{Cohen89}. In an oblivious bit-fixing source $X$ of length $n$,
$k$ bits in $X$ are unbiased and independent, and the remaining $n-k$ bits are fixed. The positions
of the $k$ bits are unknown. In fact,
oblivious bit-fixing sources is a special type of independent sources that we studied in the previous sections, where
all the bits in the source are independent of each other, among them, $k$ bits have probability $1/2$ and
the other $n-k$ bits have  probability either $0$ or $1$.  So our conclusions about the application of sparse random matrices on
independent sources still can work here.

Another type of bit-fixing sources are nonoblivious. Unlike the oblivious case, in nonoblivious bit-fixing sources,
the remaining $n-k$ bits are linearly determined by the $k$ independent and unbiased bits. Such sources were originally studied in the context of
collective coin flipping \cite{Ben90}.

Generally, we can describe a (nonoblivious) bit-fixing source in the following way:
Let $Z\in \{0,1\}^k$ be an independent and unbiased sequence, the source $X\in \{0,1\}^n$ can be written as
$X=ZA$, where $A$ is an unknown $k\times n$ binary matrix such that there are $k$ columns in $A$ that form
an identity matrix.

\begin{Example} One example of such a matrix $A$ is
$$A=\left(
      \begin{array}{ccccc}
        0 & 1 & 0 & 0 & 1 \\
        1 & 0 & 0 & 1 & 0 \\
        1 & 0 & 1 & 0 & 1 \\
      \end{array}
    \right).
$$
If we consider the columns $2,4,3$, then they form an identity matrix.
\end{Example}

Given a bit-fixing source with $k$ independent and unbiased bits, one cannot extract more than $k$ random bits that are arbitrarily close to truly random bits. That's because the entropy of the output sequence must be upper bounded by the entropy of the input sequence, which is $k$.

\begin{Lemma}  \label{lin_theorem6}
Let $X=x_1x_2...x_n\in \{0,1\}^n$ be a bit-fixing source in which
$k$ bits are unbiased and independent.
Let $M$ be an $n\times m$ uniform random matrix such that each entry of $M$ is $0$ or $1$ with probability $\frac{1}{2}$.
Given $Y=XM$, then we have
$$P_M[\rho(Y)\neq 0]\leq 2^{m-k}.$$
\end{Lemma}

\proof For a bit-fixing source $X\in\{0,1\}^n$, we can write it as $X=ZA$, where $Z\in \{0,1\}^k$ is an independent and unbiased sequence. Hence,
$$Y=XM=ZAM=ZB,$$
in which $B=AM$ is an $k\times m$ matrix.

We see that all the columns of $B$ are independent of each other because the $i$th column of $B$ only depends
on the $i$th column of $M$ for all $1\leq i\leq m$. Furthermore, it can be proved that
each column of $B$ is a vector in which all the elements are independent of each other and each element is $0$ or $1$ with probability $1/2$.
To see this, we consider an entry in $B$, which is $B_{ij}=\sum_{k}A_{ik}M_{kj}$.
Given this $i$, according to the definition of $A$, we can always find a column $r$ such that
only the element in the $i$th row is $1$ and all the other elements in this column are $0$s.
So we can write $$B_{ij}=M_{ir}+\sum_{k\neq r}A_{ik}M_{kj},$$
$$B_{i'j}=\sum_{k\neq r}A_{i'k}M_{kj}, \textrm{ for } i'\neq i,$$
where $M_{ir}$ is an unbiased random bit independent of $M_{kj}$ with $k\neq r$.  In this case, $B_{ij}$ is independent
of $B_{i'j}$ with $i'\neq i$. Hence, we can conclude that $B$ is a random matrix in which each entry is $0$ or $1$ with probability $1/2$.

According to Lemma \ref{lemma3}, we get that $\rho(Y)=0$ if and only if
$ZBu^T$ is an unbiased random bit for all $u\neq 0$.

Hence,
\begin{eqnarray}
  P_M[\rho(Y)\neq 0] &\leq &  \sum_{u\neq 0} P_M[ZBu^T \textrm{ is fixed }] \nonumber\\
  &=& \sum_{u\neq 0} P_B[Bu^T= 0],\label{equ_fix_1}
\end{eqnarray}
where $Bu^T$ is a random vector with each element being $0$ or $1$ with
probability $1/2$ for all $u\neq 0$.
So $$P_B[Bu^T= 0]=2^{-k}.$$

Finally, we can get that
$$P_M[\rho(Y)\neq 0]\leq \sum_{u\neq 0} 2^{-k} \leq 2^{m-k}.$$
This completes the proof.
\hfill\QED

According to the above lemma, by using a uniform random matrix with $m-k\leq 0$,
we can generate an independent and unbiased sequence from a bit-fixing source with almost probability $1$. In the following theorem,
we show that sparse random matrices can also work for bit-fixing sources.

\vspace{0.1in}\hspace{-0.15in}\textbf{Theorem \ref{theorem3_1}.} \emph{Let $X=x_1x_2...x_n\in \{0,1\}^n$ be a bit-fixing source in which
$k$ bits are unbiased and independent.  Let $M$ be an $n\times m$ binary random matrix in which the probability for each entry being $1$
is $p= w(\frac{\log n}{n})\leq \frac{1}{2}$.
Assume $Y=XM$.
If $\frac{m}{k}<1$, as $n$ becomes large enough, we have that $\rho(Y)=0$ with almost probability $1$, i.e.,
$$P_M[\rho(Y)= 0]\rightarrow 1.$$}
\vspace{-0.15in}

\proof
According to  Equ. (\ref{equ_fix_1}), we have
\begin{eqnarray*}
P_M[\rho(Y)\neq 0] &=&\sum_{u\neq 0} P_M[AMu^T=0].
\end{eqnarray*}

When $u\neq 0$, $Mu^T$ is a random vector in which all the elements are independent of each other.
Let $|u|=j$, then according to Lemma \ref{lemma1}, the probability for each element in
 $Mu^T$ being $1$ is
 $$p_j\in [\frac{1}{2}(1-(1-2p)^j),\frac{1}{2}(1+(1-2p)^j)].$$

Let $v^T=AMu^T$ and use $v_i^T$ denote its $i$th element, then
$$P_M[v^T=0]=\prod_{i=1}^k P[v_i^T=0|v_1^T=0,...,v_{i-1}^T=0].$$

According to the constraint on $A$, we know that there exists a column that is $[0,...,0,1,0,...,0]^T$, in which
only the entry in the $i$th row is $1$. Without loss of generality, we assume that this column is the $r$th column.
Then we can write
$$v_i^T=(Mu^T)_r+ \sum_{t\neq r, t=1}^{n}a_{it}(Mu^T)_t,$$
where $(Mu^T)_r$ is $1$ with probability $p_j\in [\frac{1}{2}(1-(1-2p)^j),\frac{1}{2}(1+(1-2p)^j)]$, and it is independent of $v_1^T, v_2^T, ...$
Hence,
\begin{eqnarray*}
&&P_M[v_i^T=0|v_1^T=0,...,v_{i-1}^T=0]\\
&=& \sum_{a=0}^1 P_M[(Mu^T)_r=a] \\
&& \quad\times P_M[\sum_{t\neq r, t=1}^{n}a_{it}(Mu^T)_t=a|v_1^T=0,...,v_{i-1}^T=0]\\
&\leq & \max_{a=0}^1 P_M[(Mu^T)_r=a]\\
&=& \frac{1}{2}(1+(1-2p)^j).
\end{eqnarray*}

So when $|u|=j$, we can get
$$P_M[AMu^T=0]\leq (\frac{1}{2}(1+(1-2p)^j)^k.$$

As a result,
$$P_M[\rho(Y)\neq 0]\leq \sum_{j=1}^m \nchoosek{m}{j} (\frac{1}{2}(1+(1-2p)^j))^k.$$

Let us divide it into two parts,
$$\gamma_1= \sum_{j=1}^{\frac{\log\frac{1}{\epsilon}}{2p}}\nchoosek{m}{j} (\frac{1}{2}(1+(1-2p)^j))^k,$$
$$\gamma_2= \sum_{j=\frac{\log\frac{1}{\epsilon}}{2p}}^{m}\nchoosek{m}{j} (\frac{1}{2}(1+(1-2p)^j))^k,$$
where $\epsilon$ is arbitrary small. Then
$$P_M[\rho(Y)\neq 0]\leq \gamma_1+\gamma_2.$$

According to Lemma \ref{lemma4}, we can get that the first part
$ \gamma_1\rightarrow 0$
as $n\rightarrow 0$.

For the second part $\gamma_2$, it is easy to show that for any $\epsilon>0$, when $n$ (or $k$) is large enough
\begin{eqnarray*}
\gamma_2&=& \sum_{j=\frac{\log\frac{1}{\epsilon}}{2p}}^{m}\nchoosek{m}{j} (\frac{1}{2}(1+(1-2p)^j))^k\\
&\leq & \sum_{j=\frac{\log\frac{1}{\epsilon}}{p}}^{m}\nchoosek{m}{j} (\frac{1}{2}(1+\epsilon))^k\\
&\leq & 2^{m-k}(1+\epsilon)^k.
\end{eqnarray*}

As a result, if $m-k\log \frac{2}{1+\epsilon}\ll 0$ for an arbitrary $\epsilon$, then
$P_M[\rho(Y)\neq 0]$ can be very small. Therefore, we get the conclusion in the theorem.

This completes the proof.
\hfill\QED

We see that sparse random matrices are asymptotically optimal for extracting randomness from bit-fixing sources. Now a question is
whether we can find an explicit construction of linear transformations for extracting randomness efficiently from any bit-fixing source specified by $n$ and $k$.
Unfortunately, the answer is negative. The reason is that in order to extract independent random bits, it requires $XMu^T$ to be an unbiased random bit for all $u\neq 0$ (See the proof above). So $\|Mu^T\|>n-k$ for all $u\neq 0$, otherwise we are able to find a bit-fixing source $X$ such that $XMu^T$ is fixed. Such a bit-fixing source can be constructed as follows: Assume
 $X=x_1x_2...x_n$, if $(Mu^T)_i=1$ we set $x_i$ as an unbiased random bit, otherwise we set $x_i=0$ being fixed.
It further implies that if we have a linear code with generator matrix $M^T$, then its minimum distance should be more than $n-k$. But for such a matrix, its efficiency ($\frac{m}{n}$)
 is usually very low. For example, when $k=\frac{n}{2}$, we have to find a linear code with minimum distance more than $\frac{n}{2}$. In this case, the dimension of the code, i.e., $m$,  is much smaller than $k$, implying a low efficiency in randomness extraction.

\section{Linear-Subspace Sources}
\label{lin_section_subspace}

In this previous section, we studied a bit-fixing source $X\in \{0,1\}^n$, which can be written as $ZA$, where $Z\in \{0,1\}^k$ is an independent and unbiased sequence and $A$ is an unknown $k\times n$ matrix that embeds an identity matrix.
Actually, we can generalize the model of bit-fixing sources in two directions.
First, the matrix $A$ can be generalized to any full-rank matrix.
Second, the sequence $Z$ is not necessary being independent and unbiased. Instead, it can be any random source described in this paper,
like an independent source or a hidden Markov source. The new generalized source $X$ can be treated as a mapping of another source $Z$
into a linear subspace of higher dimensions, so we call it a linear-subspace source.
The rows of the matrix $A$, which are independent of each other, form the basis of the linear subspace.
Linear-subspace sources are good descriptions of many natural sources, like sparse images studied in compressive sensing.

First, let us consider the case that the matrix $A$ is an arbitrary unknown full rank matrix and $Z$ is still an independent and unbiased sequence.
\begin{Lemma}\label{lin_theorem7}
Let $X=x_1x_2...x_n\in \{0,1\}^n$ be a source such that $X=ZA$ in which $Z$ is an independent and unbiased sequence, and $A$ is an unknown $k \times n$ full-rank matrix. Let $M$ be an $n\times m$ random matrix such that each entry of $M$
is $1$ with probability $p= w(\frac{\log n}{n})\leq \frac{1}{2}$.
Assume $Y=XM$.
If $\frac{m}{k}<1$, as $n$ becomes large enough, we have $\rho(Y)=0$ with almost probability $1$, i.e.,
$$P_M[\rho(Y)= 0]\rightarrow 1.$$
\end{Lemma}

\proof In the proof of Theorem \ref{theorem3_1}, we have
\begin{eqnarray*}
P_M[\rho(Y)\neq 0] &=&\sum_{u\neq 0} P_M[AMu^T=0].
\end{eqnarray*}

If the matrix $A$ has full rank, than we can write
$$A=UR,$$
where $det(U)\neq 0$ and $R$ is in row echelon form. We see that
$RZ$ is a nonoblivious bit-fixing source.

Since $det(U)\neq0$, $AMu^T=0$ is equivalent to $RMu^T=0$. Therefore, \begin{eqnarray*}
P_M[\rho(Y)\neq 0] &=&\sum_{u\neq 0} P_M[RMu^T=0].
\end{eqnarray*}

Based on the proof of Theorem \ref{theorem3_1}, we can get the conclusion in the lemma.

This completes the proof.
\hfill\QED

Furthermore, we generalize the sequence $Z$ to a general independent source in which
the probability of each bit is unknown and the min-entropy of the source is $H_{\min}(Z)$.

\vspace{0.1in}
\hspace{-0.15in}\textbf{Theorem \ref{theorem4_1}.} \emph{
Let $X=x_1x_2...x_n\in \{0,1\}^n$ be a source such that $X=ZA$ in which $Z$ is an independent sequence and $A$ is an unknown $k \times n$ full-rank matrix. Let $M$ be an $n\times m$ random matrix such that each entry of $M$
is $1$ with probability $p= w(\frac{\log n}{n})\leq \frac{1}{2}$.
Assume $Y=XM$.
If $\frac{m}{H_{\min}(X)}<1$, as $n\rightarrow\infty$,
$\rho(Y)$ converges to $0$ in probability, i.e.,
$$\rho(Y) {\quad \buildrel p \over \rightarrow \quad } 0.$$
}
\vspace{-0.15in}

\proof Let $\delta_i$ be the bias of $z_i$ in $Z$ for all $1\leq i\leq k$. According to Equ.~(\ref{lin_equ_coin_2}),
we can get
$$E_M[\rho(Y)]\leq \frac{1}{2} \sum_{u\neq 0} \sum_{v\in\{0,1\}^k} P_M[AMu^T=v^T] \prod_{i=1}^k (2\delta_i)^{v_i}.$$

When $\|u\|=j$, $Mu^T$ is an independent sequence in which each bit is one with probability $$p_j\in [\frac{1}{2}(1-(1-2p)^j),\frac{1}{2}(1+(1-2p)^j)].$$

In Theorem \ref{lin_theorem7}, we have proved that
$$P_M[AMu^T=0]\leq (\frac{1}{2}(1+(1-2p)^j)^k.$$

Using a same idea, if $A=UR$ with $det(U)\neq 0$ and $R$ in row echelon form, we can write
\begin{eqnarray*}
&&P_M[AMu^T=v^T]\\
&=&P_M[RMu^T=U^{-1}v^T]\\
&=&\prod_{i=1}^k P_M[(RMu^T)_i=(U^{-1}v^T)_i\\
&&\quad\quad| (RMu^T)_{i-1}=(U^{-1}v^T)_{i-1},...]\\
&\leq & (\frac{1}{2}(1+(1-2p)^j)^k
\end{eqnarray*}
for all $v^T\in \{0,1\}^k$.

Hence
$$E_M[\rho(Y)]\leq \frac{1}{2} \sum_{j=1}^m {\nchoosek{m}{j}} (\frac{1}{2}(1+(1-2p)^j)^k  \prod_{i=1}^k (1+2\delta_i).$$

In the next step, following the proof of Theorem \ref{theorem1_1}, we can get
that if $\frac{m}{H_{\min}(Z)}<1$, as $n\rightarrow \infty$,
$$E_M[\rho(Y)]\rightarrow 0.$$
It is equivalent to $\rho(Y) {\quad \buildrel p \over \rightarrow \quad } 0.$

Since $H_{\min}(Z)=H_{\min}(X)$, we can get the conclusion in the theorem.

This completes the proof.
\hfill\QED

A similar result holds if $Z$ is a hidden Markov sequence. In this case, we have the following theorem.

\begin{Theorem}
Let $X=x_1x_2...x_n\in \{0,1\}^n$ be a source such that $X=ZA$ in which $Z\in \{0,1\}^k$ is a hidden Markov sequence described in Section \ref{lin_section_Markov}, and $A$ is an unknown $k \times n$ full-rank matrix.
Let $M$ be an $n\times m$ random matrix such that each entry of $M$
is $1$ with probability $p= w(\frac{\log n}{n})\leq \frac{1}{2}$.
Assume $Y=XM$.
If $\frac{m}{k\log_2\frac{2}{1+\sqrt{e}}}<1$, as $n\rightarrow\infty$,
$\rho(Y)$ converges to $0$ in probability, i.e.,
$$\rho(Y) {\quad \buildrel p \over \rightarrow \quad } 0.$$
\end{Theorem}

From the above theorems, we see that by multiplying an invertible matrix to a given source does not affect the extracting capability of sparse random matrices.

\section{Implementation for High-Speed Applications}
\label{lin_section_stream}

In this section, we discuss the implementation of linear transformations in high-speed random number generators, where
the physical sources usually provide a stream rather than a sequence of finite length.
To generate random bits, we can apply a linear transformation to the incoming stream based on block by block, namely, we divide the incoming stream into blocks and
generate random bits from each block separately. Such an operation can be finished by software or hardware like FPGAs \cite{ElKurdi08,Zhuo05}.

Another way is that we process each bit when it arrives. In this case, let $M=\{m_{ij}\}$ be
an $n\times m$ matrix (such as a sparse random matrix) for processing the incoming stream and let $V\in \{0,1\}^m$ denote
a vector that stores $m$ bits. The vector $V$ is updated dynamically in response of the incoming bits.
When the $i$th bit of the stream, denoted by $x_i$, arrives we
do the following operation on $V$,
$$V\rightarrow V+x_i M_{1+(i\textrm{ mod }n)},$$
where $M_j$ is the $j$th row in the matrix $M$.
Specifically, we can write the vector $V$ at time $i$ as $V[i]$ and denote its $j$th element as $V_j[i]$.
To generate (almost) random bits, we output the bits in $V$ sequentially and cyclically with a lower rate than that of the incoming stream. Namely,
we generate an output stream $Y=y_1y_2...$ such that
$$y_i=V_{1+(i\textrm{ mod }m)}[n+\lfloor\frac{ni}{m}\rfloor].$$
So the rate of the output stream is $\frac{m}{n}$ of the incoming stream.
In this method, the expected computational time for processing a single incoming bit is proportional to
the number of ones in $M$ over $n$. According to our results of sparse random matrices,
it can be as low as $(\log n)^\alpha$ with any $\alpha>1$ asymptotically. So this method is
computationally very efficient, and the working load is well balanced.

\begin{table*}
  \centering
   \caption{Asymptotical efficiencies of linear transformations for extracting randomness from different sources}
\renewcommand{\arraystretch}{1.5}
  \begin{tabular}{|p{2cm}|p{6cm}|p{6cm}|}
    \hline
    Source $X=x_1x_2...x_n$ &Sparse Random Matrices & Generator Matrices \\
    \hline
    Independent Sources  & $H_{\min}(X)$ &
    $n\log_2\frac{2}{1+e}$ \par if $P[x_i=1]\in [\frac{1}{2}-\frac{e}{2},\frac{1}{2}+\frac{e}{2}]$  \\
    \hline
    Hidden Markov Sources & $n\log_2\frac{2}{1+\sqrt{e}}$ \par if $P[x_{i_2}=1|\theta_{i_1},\theta_{i_3}]\in [\frac{1}{2}-\frac{e}{2},\frac{1}{2}+\frac{e}{2}]$ &$n\log_2\frac{2}{1+\sqrt{e}}$ \par if $P[x_{i_2}=1|\theta_{i_1},\theta_{i_3}]\in [\frac{1}{2}-\frac{e}{2},\frac{1}{2}+\frac{e}{2}]$ \\
    \hline
    Bit-Fixing Sources  & $H_{\min}(X)$ & NA\\
    \hline
    Linear-Subspace Sources & $H_{\min}(X)$ \par if $X=AZ$ with $A$ full-rank and $Z$ independent & NA\\
    \hline
  \end{tabular}
 \label{lin_table1}
\end{table*}

\section{Conclusion}
\label{lin_section_conclusion}

In this paper, we demonstrated the power of linear transformations in randomness extraction from a few types of weak random sources, including
independent sources, hidden Markov sources, bit-fixing sources, and linear-subspace sources, as summarized in Table \ref{lin_table1}.
Compared to the existing methods, the constructions of linear transformations are much simpler, and they can be easily implemented using FPGAs;
these properties make methods based on linear transformations very practical. To reduce the hardware/computational complexity, we prefer sparse matrices rather than high-density matrices, and we proved that sparse random matrices can work as well as uniform random matrices.
Explicit constructions of efficient sparse matrices remain a topic for future research.

\end{document}